\documentclass[aps,reprint, onecolumn,superscriptaddress, prl]{revtex4-1}
\usepackage{graphicx}
\usepackage{amsmath}
\usepackage{amsthm}
\usepackage{amssymb}
\usepackage{amsfonts}
\usepackage{siunitx}

\usepackage{blindtext}
\usepackage{comment}
\usepackage{xcolor}
\usepackage{xr}
\usepackage{lineno}

\begin{document}
\title{Anomalous Angiogenesis in Retina}

\author{Roc\'{\i}o Vega}
\affiliation{Gregorio Mill\'an Institute, 
 Fluid Dynamics, Nanoscience and Industrial Mathematics, 
 and Department of Mathematics,
 Universidad Carlos III de Madrid, 28911 Legan\'es, Spain} 
 \author{Manuel Carretero}
\affiliation{Gregorio Mill\'an Institute, 
 Fluid Dynamics, Nanoscience and Industrial Mathematics, 
 and Department of Mathematics,
 Universidad Carlos III de Madrid, 28911 Legan\'es, Spain}
 \author{Luis L. Bonilla}\affiliation{Gregorio Mill\'an Institute, 
 Fluid Dynamics, Nanoscience and Industrial Mathematics, 
 and Department of Mathematics,
 Universidad Carlos III de Madrid, 28911 Legan\'es, Spain}


\begin{abstract}
Age-related macular degeneration (AMD) may cause severe loss of vision or blindness particularly in elderly people. Exudative AMD is characterized by angiogenesis of blood vessels growing from underneath the macula, crossing the blood-retina barrier (that comprise Bruch's membrane, BM, and the retinal pigmentation epithelium RPE), leaking blood and fluid into the retina and knocking off photoreceptors. Here, we simulate a computational model of angiogenesis from the choroid blood vessels via a cellular Potts model, as well as BM, RPE cells, drusen deposits and photoreceptors. Our results indicate that improving AMD may require fixing the impaired lateral adhesion between RPE cells and with BM, as well as diminishing Vessel Endothelial Growth Factor  (VEGF) and Jagged proteins that affect the Notch signaling pathway. Our numerical simulations suggest that anti-VEGF and anti-Jagged therapies could temporarily halt exudative AMD while addressing impaired cellular adhesion could be more effective on a longer time span.
\end{abstract}

\maketitle

\section{Introduction}
\label{sec:intro}
Among diseases that cause disability but not substantial mortality, age-related macular degeneration (AMD) may cause severe loss of vision or blindness in many people, particularly the elderly. Wong {\em et al} have projected that 196 million people will be affected by age-related macular degeneration in 2020, increasing to 288 million by 2040 \cite{won14}, which is likely an underestimation \cite{jon14}. Exudative or wet AMD is characterized by a breakdown of the blood-retina barrier: blood vessels grow from underneath the macula and leak blood and fluid into the retina \cite{jag08,niv14}. These blood vessels and their leaking may form scars leading to permanent loss of central vision. Diagnosis of wet AMD \cite{niv14} has improved with important non invasive techniques such as optical coherence tomography \cite{man10,cos15} or, quite recently, transscleral optical phase imaging \cite{laf20}. The retina contains many membranes and tissue layers that make imaging cells and understanding pathologies difficult. The growth of blood vessels in the retina is well documented in pathological cases, such as wet AMD, and in normal cases, such as retinal vascularization in fetuses and newborns \cite{GG05,fru07,sco10,sel18}. In these cases, blood vessels grow out of a primary vessel, which is a complex multiscale process called angiogenesis. Retinal angiogenesis adds a complex geometry to the process.

Normal angiogenesis determines organ growth and regeneration, wound healing, repair of tissues, etc  \cite{GG05,fru07,sco10,sel18,car05,CT05,car11,ton00,fig08}. In these processes, tissue  inflammation may occur and cells may experience hypoxia. Then they may activate signaling pathways leading to secretion of pro-angiogenic proteins, such as Vessel Endothelial Growth Factor (VEGF). VEGF diffuses in the tissue, binds to extracellular matrix (ECM) components and forms a spatial concentration gradient in the direction of hypoxia. In retinal vascularization, astrocyte neurons issuing from the optical nerve form a network and locally secrete VEGF \cite{fru07,sco10,sel18}. VEGF molecules that reach an existing blood vessel diminish adhesion of its cells and activate the tip cell phenotype in endothelial cells (ECs) of the vessel through the Notch signaling pathway. Tip cells grow filopodia with many VEGF receptors, pull the other ECs, open a pathway in the ECM, lead the new sprouts, and migrate in the direction of increasing VEGF concentration \cite{ger03}. {\em Branching of new sprouts} occurs as a result of signaling and mechanical cues between neighboring ECs \cite{hel07,jol15,pag19,veg20}. ECs in growing sprouts alter their shape to form a lumen connected to the initial vessel that is capable of carrying blood \cite{geb16}. Sprouts meet and merge in a process called {\em anastomosis} to improve blood circulation inside the new vessels. Poorly perfused vessels may become thinner and their ECs, in a process that inverts angiogenesis, may retract to neighboring vessels leading to a more robust blood circulation \cite{fra15}. Thus, the vascular plexus remodels into a highly organized and hierarchical network of larger vessels ramifying into smaller ones \cite{szy18}. In normal processes of wound healing or organ growth, the cells inhibit the production of growth factors when the process is finished.

\begin{figure}[h]
	\begin{center}
		\includegraphics[width=0.85\linewidth]{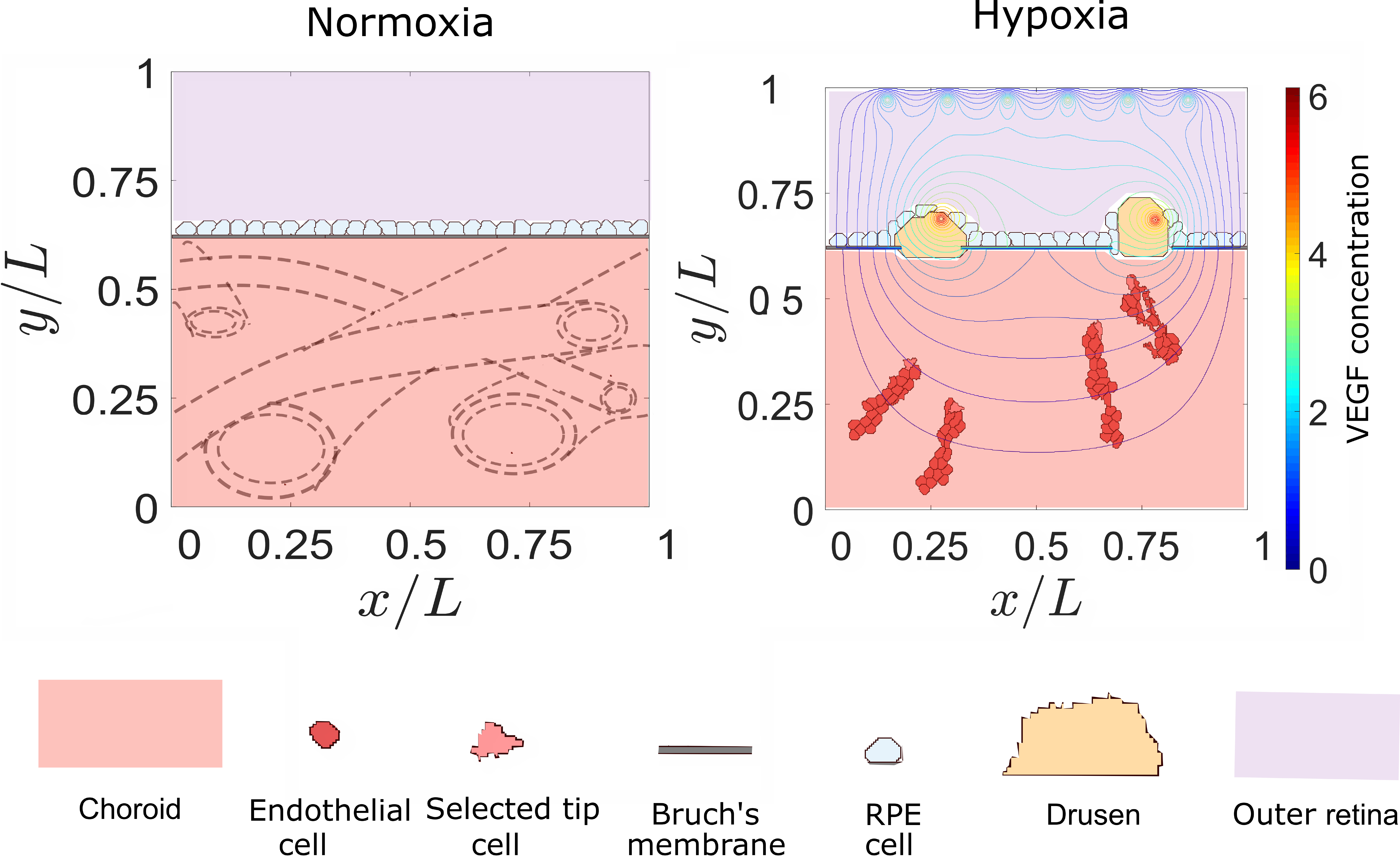}
\end{center}
\vskip-3mm 
\caption{Sketch of a two dimensional section of the macula, including choroid and choroid blood vessels, Bruch's membrane, retinal pigment cells, drusen, outer retina, epithelial cells, and tip cells.  Left panel: normal supply of oxygen. Right panel: hypoxia and incipient choroid neovascularization. \label{fig1}}
\end{figure}

Pathological angiogenesis changes the previous picture in important ways. For example, hypoxic tumor cells produce VEGF that induces angiogenesis, interfere with Notch signaling and promotes tumor growth \cite{duf08}. In {\em wet AMD}, negative control mechanisms are breached,  angiogenic sprouts issue from choroid  blood vessels, cross Bruch's membrane (BM), pass retinal pigment epithelium (RPE) cells and knock down photoreceptors producing loss of central vision \cite{sel18}. Under normal circumstances, oxygen and nutrients diffuse from the choroid vascular layer through the thin BM and the layer of RPE cells to reach photoreceptors. Inversely, RPE process photoreceptor debris, including shed photoreceptor outer segments, pass it through BM, and the debris is removed by the choriocapillaries in the choroid; see the left panel of Fig.~\ref{fig1}. With age, BM becomes thicker and a variety of factors such as oxidative stress, immune activation, genetic constitution and local anatomy of the neural retina, RPE and BM may affect the ability of the latter to process debris. Lipid deposits called {\em drusen} may form on different layers of BM and the RPE \cite{boo10}. Calcification of BM also diminishes the adhesion of RPE cells. Damage to RPE cells and their loss is compensated by the same wound healing mechanism as in other epithelial cells. However, this process is impaired in AMD tissues \cite{boo10}. Drusen deposits thicken BM and create a barrier that interferes with the normal situation. They decrease the diffusion of oxygen and nutrients from the choroid to the photoreceptors and the RPE in one direction, and decrease the removal of debris by the choriocapillaries into the other direction \cite{mam20}. While this {\em dry AMD} produces some loss of vision (severe in the case of geographical atrophy of RPE \cite{geh06}), it also leads to relative hypoxia within retinal layers. In response, the RPE secrete proangiogenic factors, such as VEGF and vasculogenic and inflammatory cytokines, that go into the choroidal space. This may start angiogenesis thereby producing {\em choroidal neovascularization} (CNV); see the right panel of Fig.~\ref{fig1}. CNV vessel sprouts may penetrate BM and remain underneath the RPE layer (type 1 or `occult' CNV), or surpass it and go through the outer retina (type 2 or `classic' CNV). In the later condition, sprouts may leak blood and fluid that eventually produce scars and the death of photoreceptors, which signals the wet phase of AMD \cite{geh06}; see also sketch in Fig.~1 of Ref.~\cite{mam20}. During AMD, RPE cells degenerate, lose their characteristic epithelial morphology and function, enabling their migration into the retina and the sub-RPE space \cite{rad15}. These cells may have undergone a reversible Epithelial-to-Mesenchymal Transition (EMT), detach themselves from the RPE and migrate to survive the adverse environment during AMD \cite{shu20}. The loss of RPE cells due to the EMT and the similar transformation of endothelial cells to a motile phenotype during angiogenesis are very important factors in wet AMD \cite{shu20}. In this disease, the process of RPE detachment and dissociation is crucial, and the initiation of the EMT requires the disruption of RPE cell-cell contact \cite{tam10}.

We have just sketched the complexity of the retinal structures involved in AMD. While there are many mathematical and computational models describing the development of retinal vascularization, less work has been devoted to such models for AMD; see e.g., \cite{rob16}. Early modeling research on CNV focus on relating the blood flow in the CNV to that in the underlying choriocapillaries  in an appropriate two dimensional (2D) geometry \cite{flo01}. Darcy's law for a porous medium is used to model flow in choriocapillaries and in blood vessels that connect them to the CNV. Changes in the flow in the connecting vessels strongly influence the flow through the CNV and controling the connecting vessels may be used to block the flow in the CNV, with beneficial effects for wet AMD \cite{flo01}. A similar idea but using the incompressible Navier-Stokes equation has been recently used to study drug delivery across the blood-retina barrier \cite{dav20}. These studies consider fixed choroid and CNV networks, ignoring the development and progression of the latter. To account for the formation and expansion of the CNV, Shirinifard {\em et al} have used a 3D cellular Potts model of the choroid and outer retina \cite{shi12}. 
They conclude that failures in cellular adhesion determine the formation and expansion of CNV. Cells include tip and stalk ECs, cells at RPE and BM, photoreceptor cell outer and inner segments, ECM and fluid regions. Continuum equations model media surrounding cells, VEGF, oxygen and matrix metalloproteinases and are coupled to the Potts hamiltonian that is updated using a modified Metropolis algorithm. EC chemotaxis and haptotaxis are implemented. Stalk cells may increase their volume and proliferate, BM cells may decrease their volume and die. Different cells have different adhesion parameters. The model does not include blood flow, signaling pathways and change of EC phenotype, or drusen \cite{shi12}.

Here we use a 2D Potts model that includes EC Notch signaling, chemotaxis, haptotaxis and durotaxis \cite{veg20} to ascertain the influence of these mechanisms on AMD. We consider the simple geometry sketched in Fig.~\ref{fig1}: a square domain in which BM separates the choroid crisscrossed by blood vessels, which may issue angiogenic sprouts, from RPE cells, eventual drusen and a subretinal space on top of which there are photoreceptors. The choroid vessels may issue sprouts at randomly chosen points provided the VEGF concentration surpasses some threshold in those points. The growth of drusen above RPE cells turns on VEGF sources that attract the sprouts issued from the choroid vessels to them. Once ECs have crossed BM, they either form subRPE type 1 CNV or subretinal type 2 CNV. Type 1 CNV occurs if the sprouts form a network between BM and the RPE cells, whereas type 2 CNV occurs if the sprouts succeed moving beyond the RPE layer and towards the VEGF emitting photoreceptors. We find that adhesion between RPE cells, between ECs, and between RPE cells and BM decides whether angiogenic sprouts succeed in invading the subRPE space or the subretinal space, thereby producing type 1 or 2 CNV, respectively. We study how local VEGF gradients and Notch signaling proteins affect CNV in presence of drusen and defects in BM. Notch signaling dynamics confirms that CNV is an example of pathological angiogenesis with thin and leaky capillary sprouts \cite{veg20}.

\section{Materials and Methods} 
\label{sec:model}
To describe angiogenesis in the retina, we need a model able to describe cellular processes at cellular and subcellular sizes. The cellular Potts model (CPM) \cite{gra92} is particularly useful at these scales, as it incorporates in a natural way constraints for the volume, area or length of the cells, as well as adhesion between cells or with the extracellular matrix (haptotaxis) \cite{shi12,gra92}.  Attraction due to chemical gradients (chemotaxis) \cite{bau07} or to substrate stiffness gradients  (durotaxis)\cite{oer14} have also been added to CPMs. Strains in the cells together with the unsupervised K-means algorithm can be used to implement branching of growing sprouts \cite{veg20}. The phenotype of leading tip cells or follower stalk cells is  decided by the Notch signaling pathway, and the corresponding dynamics \cite{boa15} can also be incorporated to the CPM \cite{veg20}. 

\subsection{Cellular Potts model}
In our simulations, we consider different entities: the choroid, Bruch's membrane, Retinal Pigmented Epithelium cells, endothelial cells, extracellular matrix, photoreceptors, and drusen. We ignore the outer segments of photoreceptors and their dynamics. Thus, there is a free space between the RPE layer and the photoreceptors. We fix the  number of drusen, and of RPE cells, whereas the number of ECs varies. Different cells comprise a number of elementary squares or pixels in a square domain $\Omega$ of side $L$ (in numerical simulations, $L= 400\ \mu$m). See Appendix \ref{app1} for the precise labels (or spins) of pixels belonging to different cells. For each pixel configuration, we define the Hamiltonian 
\begin{eqnarray}
H\!&=&\! \displaystyle\sum_{\sigma} \rho_{\text{area}} \left(\frac{a_{\sigma} - A_\sigma}{A_\sigma}\right)^2 + \displaystyle\sum_{\sigma}\rho_{\text{perimeter}} \left(\frac{p_{\sigma} - P_\sigma}{P_\sigma}\right)^2 + \displaystyle\sum_{\sigma} \rho_{\text{length}} \left(\frac{l_{\sigma} - L_\sigma}{L_\sigma}\right)^2 \nonumber\\
\!&+&\! \displaystyle\sum_{\mathbf{x}, \mathbf{x}' \in \Omega} \rho_\text{adh}^{\Sigma_\sigma, \Sigma_{\sigma'}} \left(1 - \delta_{\sigma,\sigma'}\right)\! + H_{\text{durot}} + H_{\text{chem}}, \label{eq1}
\end{eqnarray}
where the three first terms are sums over cells. These terms impel them to reach target areas, perimeters and lengths with strengths given by their Potts parameters $\rho_\text{area}$, $\rho_\text{perimeter}$ and $\rho_\text{length}$. Numerical values of target areas $A_\sigma$, perimeters $P_\sigma$, and lengths  $L_\sigma$ can be found in Table \ref{t0}. The fourth term (haptotaxis) sums over all pixels and accounts for adhesion between elements. It is zero for pixels belonging to the same cell and calibrates the repulsion between pixels belonging to different cells (adhesion is stronger for smaller repulsion), depending on the value of the corresponding Potts parameter $ \rho_\text{adh}^{\Sigma_\sigma, \Sigma_{\sigma'}}$. The fifth and sixth terms correspond to durotaxis and chemotaxis, impelling cells to move toward gradients of stiffness and VEGF concentration, respectively \cite{veg20}. See also Appendix \ref{app1}. At each Monte Carlo time step (MCTS) $t$, we select randomly a pixel $\mathbf{x}$, belonging to object $\Sigma_\sigma$, and propose to copy its spin $\sigma(\mathbf{x})$ to a neighboring (target) pixel $\mathbf{x'}$ that does not belong to $\Sigma_{\sigma(\mathbf{x})}$. The proposed change in the spin configuration (spin flip) changes the configuration energy by an amount $\triangle H|_{\sigma(\mathbf{x})\to\sigma(\mathbf{x'})}$, and it is accepted with probability $P\left(\sigma(\mathbf{x}) \rightarrow \sigma(\mathbf{x}')\right)(t) =\{e^{-\Delta H|_{\sigma(\mathbf{x}) \rightarrow \sigma(\mathbf{x}')}/T} \mbox{if }\Delta H>0, \mbox{and } 1 \mbox{ if }\Delta H\leq 0\}$ (Metropolis algorithm) \cite{gra92,oer14}. BM does not change throughout the simulation. Thus, Monte Carlo (MC) attempts involving $\mathbf{x} \in \Sigma_\text{BM}$ or $\mathbf{x'}\in \Sigma_\text{BM}$ are discarded. An appropriate temperature for our simulations is $T=4$.

\begin{table}[h]
\begin{center}
	\begin{tabular}{r|c|c|c|c|c|c|c}
		\text{Param.} & $A_\text{EC}$ &  $P_\text{EC}$ & $L_\text{EC}$ & $A_\text{RPE}$ & $P_\text{RPE}$ &  $A_\text{druse}$ & $P_\text{druse}$\\ \hline
		\text{Value}	&  78 $\mu$m$^2$    & 50 $\mu$m  & 60 $\mu$m & 169 $\mu$m$^2$ & 52 $\mu$m & 2827 $\mu$m$^2$ & 188  $\mu$m
	\end{tabular}
	\caption{Target areas, perimeters and length.}
	\label{t0} \end{center}
\end{table}

\subsection{Continuum fields at the extracellular scale}
\paragraph{VEGF concentration.} The VEGF concentration $C(x,y,t)$ obeys the following initial-boundary value problem \cite{bau07}:
\begin{subequations}\label{eq2}
\begin{eqnarray}
\frac{\partial C}{\partial t} = D_f \left(\frac{\partial^2 C}{\partial x^2} + \frac{\partial^2 C}{\partial y^2}\right) - \nu\, C - G(x,y,C) + A(x,y), \ (x,y) \in \Omega, \ t>0, \label{eq2a}\\
C(0,y,t) = 0= C (L,y,t), \ C(x,0,t) = 0 = C(x,L,t), \ (x,y) \in\partial\Omega, \ t>0, \label{eq2b}\\
\mathbf{n}\cdot\nabla C(x,y,t) = 0, \ (x,y) \in x_{\text{h}} \times \{ y_{\text{d}}, y_{\text{u}} \} \cup \{x_{\text{1l}}, x_{\text{1r}}, x_{\text{2l}}, x_{\text{2r}}\} \times y_{\text{v}}, t>0, \label{eq2c}\\
C(x,y,0) =0, \ (x,y) \in \Omega. \label{eq2d}
\end{eqnarray}
\end{subequations}
where $y_{\text{d}} = 246 \ \mu \text{m}, \  y_{\text{u}} = 248 \ \mu \text{m}, y_{\text{v}} =  [y_{\text{d}}, y_{\text{u}}], \ x_{\text{1l}} = 72 \ \mu \text{m}, \ x_{\text{1r}} = 128 \ \mu \text{m}, \ x_{\text{2l}} = 272 \ \mu \text{m}, \   x_{\text{2r}} = 328 \ \mu \text{m} , \ x_{\text{h}} = [0\ \mu \text{m},x_{\text{1l}}] \cup [x_{\text{1r}},x_{\text{2l}}] \cup [x_{\text{2r}},400\ \mu \text{m}]$ are the points in which BM is located with the corresponding holes.
In Eq.~\eqref{eq2a}, the amount of VEGF bound by an EC per unit time is
\begin{equation}
G(x,y,C) = \left\lbrace \begin{array}{rl}
\Gamma, & \text{ if } \Gamma \leq \upsilon\, C(x,y) \text{ and } (x,y) \in \Sigma_\text{EC},\\
\upsilon\, C, & \text{ if } 0 \leq\upsilon\, C(x,y) < \Gamma \text{ and } (x,y) \in \Sigma_\text{EC},\\
0, & \text{ if } (x,y) \notin\Sigma_\text{EC},
\end{array}\right.\label{eq3}
\end{equation}
where $\upsilon = 1 \ \text{h}^{-1}$, $D_f=0.036$ mm$^2$/h, $\nu=0.6498/$h, $S=5\times 10^{-7} \text{pg/}\mu\text{m}^2$ (corresponding to 50 ng/mL for a sample having a 10 $\mu$m height \cite{ari11,sug15}), and $\Gamma= 0.02 \text{ pg/}(\mu\text{m}^2 \text{ h})$ is the maximum amount of VEGF that it could be consumed by a cell per hour \cite{man04,bau07,veg20}. In eq.~\eqref{eq2a}, the VEGF source due to the hypoxia caused by drusen and photoreceptors is
\begin{eqnarray}
A(x,y) = \sum_{i = 1}^{N_{\text{drusen}}} \alpha_i \exp\!\left\{-\left[ \frac{  \left(x - x^{d_i} \right)^2}{2\sigma_x^2} + \frac{  \left(y - y^{d_i} \right)^2}{2\sigma_y^2}\right]\right\} 
+ \sum_{i = 1}^{N_{\text{photo}}} \alpha_i \exp\!\left\{-\left[ \frac{  \left(x - x^{p_i} \right)^2}{2\sigma_x^2} + \frac{  \left(y - y^{p_i} \right)^2}{2\sigma_y^2}\right]\right\}\!.\label{eq4}
\end{eqnarray}
Here the coefficient $\alpha_i$ is the amplitude, $(x^{d_i},y^{d_i})$  or $(x^{p_i},y^{p_i})$ is the center and $\sigma_x, \ \sigma_y$ are the x and y spreads of the blob, $\sigma_x = \sigma_y = 7$. After a sprout arrives nearby a drusen, the surrounding region ceases to be hypoxic, therefore the corresponding Gaussian of the first summation disappears from $A(x,y)$.
 
\paragraph{Durotaxis.} ECs generate mechanical strains in the substrate, perceive a stiffening of the substate along the strain orientation, and extend preferentially on stiffer substrate. The simulated ECs spread out on stiff matrices, contract on soft matrices, and become elongated on matrices of intermediate stiffness \cite{oer14}. Strains enter the durotaxis term in the Hamiltonian \eqref{eq1}, cf Appendix \ref{app1} and Refs.~\cite{oer14,veg20}.  

\subsection{Signaling processes and cell dynamics} 
A crucial distinction between ECs is that between tip and stalk phenotypes. Tip cells are highly motile, do not proliferate, act as leaders of angiogenic sprouts, sense chemical gradients and advance towards VEGF sources produced by hypoxic cells. Stalk cells proliferate and are less motile, often following tip cells. The tip-stalk cell phenotype is selected by the Notch signaling communication pathway, which is quantified by model differential equations explained in Appendix \ref{app3}. The unknowns in these equations are the  Notch, Delta-4, and Jagged-1 proteins in a cell, and the number Notch intracellular domain and VEGF molecules and of VEGF receptors in the cell. The phenotype of a cell is decided by whether the number of its VEGF molecules surpasses appropriate thresholds \cite{boa15}. See the precise criterion in Ref.~\cite{veg20} and in Appendix \ref{app3}. This means that stalk cells may become tip cells and vice versa. There are also hybrid stalk-tip cells that can lead thinner angiogenic sprouts \cite{boa15,veg20}.  Advancing blood vessels may undergo branching, thereby creating new sprouts, and fuse with existing vessels (anastomosis). The details are explained in Ref.~\cite{veg20} and Appendix \ref{app3}.

\subsection{Retinal configuration and onset of angiogenesis}
We consider a simplified configuration for the space (measuring about $L=400\,\mu$m) between the choroid and the photoreceptors as sketched in Fig.~\ref{fig1}. The choroid contains several layers comprising blood vessels of different sizes, including narrow capillaries. In a 2D section, choriocapillaries oriented in different directions may issue angiogenic sprouts that are then attracted towards openings in BM and the RPE layer. Instead of modeling the fixed choriocapillaries (parent vessels) issuing new blood vessels, we randomly generate a fixed number of points $N_\text{pv}$ that may initiate sprouts and establish an external VEGF activation threshold for the sprouts to start. The parent vessels are randomly placed  at the rectangle $0<x<L$, $0<y<0.3L=120\,\mu$m and the concentration of external VEGF satisfies Eqs.~\eqref{eq2}. The $2\, \mu$m  wide BM is  a segment placed at $y=246\,\mu$m \cite{man10,bai17} and it is followed by RPE cells with interspersed drusen, which have Gaussian sources of VEGF representing hypoxic areas. These sources placed at $y=249\,\mu$m are farther than 100 $\mu$m from the choriocapillaries, which is consistent with the criterion for hypoxia to occur. New sprouts grow from the initial points only if the external VEGF concentration in them is larger than a threshold. The described CPM causes the sprouts to advance toward the drusen and they may or may not pass BM and RPE attracted by the VEGF sources at the photoreceptors.   

After the CPM simulation begins, we need a criterion for RPE cells and photoreceptors to become hypoxic and issue VEGF. During the first hundreds of MCTS, RPE cells and drusen grow to acquire their target size \cite{friberg12,shi12}. Once a drusen $i$ reaches forty percent of its target size, it produces a hole in BM, the RPE cells around it become hypoxic and  start producing VEGF. The VEGF source associated with drusen $i$ is represented by a Gaussian function centered at $\left(x^{d_i}, y^{d_i} \right)$. This process also activates $N_{\text{photo}}$ sources of VEGF associated with photoreceptors equally spaced on the $x$ axis at $y=388\,\mu$m. We ignore the photoreceptors outer segments and their dynamics. As in the case of the sources associated with drusen, these VEGF sources are represented by Gaussian functions centered at $(x^{p_j},y^{p_j} = 388\, \mu$m)  \cite{poh20}. The holes divide BM into $N_{\text{drusen}} + 1$ pieces. Once VEGF sources are activated, new sprouts can start from the parent vessels at their predetermined sites if the external VEGF concentration there surpasses the activation threshold. VEGF sources stop emitting it when they are reached by ECs. 

\section{Results}
\label{sec:numerical}
Different adhesion parameters between ECs, and between RPE and BM cells characterize haptotaxis, which, together with VEGF gradients, determine the formation and type of CNV \cite{shi12}. In addition to confirming impaired lateral adhesion between BM and RPE and between RPE cells themselves as major drives of CNV, we explore how adhesion between ECs, chemotaxis and Notch signaling affect CNV. We find that Notch signaling proteins are markers of the CNV type that develops during AMD and decreasing the production of Jagged-1 may prevent type 2 CNV. 

\subsection{Impaired adhesion}
\label{subsec:adhesion}
Adhesion defects modify the pattern of choroidal neovascularization in the retina \cite{shi12}. The adhesion Potts parameter measures the energetic cost for cells to stay together: it is zero for pixels of the same cell and it is larger for pixels of different cells. The larger the Potts parameter between pixels of different cells is, the stronger these neighboring cells repel each other (thereby meaning weaker adhesion among them). Thus, impared adhesion among cells implies that the corresponding Potts parameter has increased with respect to the normal adhesion values. We now consider the effect that modifying Potts parameters for different cell types has on the formation and type of CNV. 

\subsubsection{Adhesion between RPE and BM}
\begin{figure}[h]
	\begin{center}
		\includegraphics[width=0.85\linewidth]{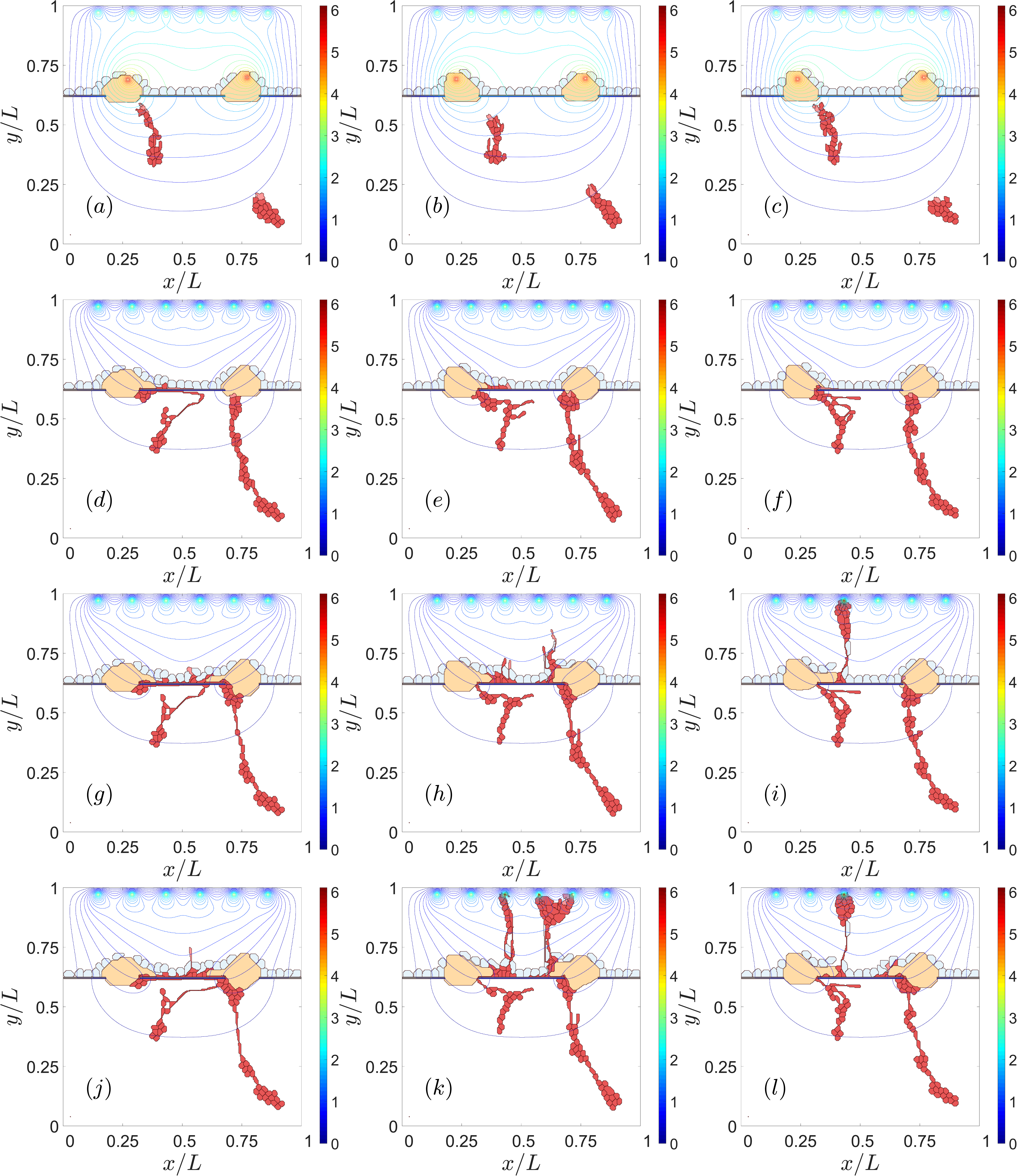}
		\caption{Effect of impaired adhesion between RPE and BM. Type 1 CNV for $\rho_\text{adh}^{\Sigma_\sigma,\Sigma_{\sigma'}}$(RPE - BM) $ = 30$, snapshots at times: (a) 601 Monte Carlo Time Steps (MCTS), (d) 1601 MCTS, (g) 4501 MCTS, (j) 9001 MCTS. Type 2 CNV for $\rho_\text{adh}^{\Sigma_\sigma, \Sigma_{\sigma'}}$(RPE - BM) $ = 6$, snapshots at times: (b) 601 MCTS, (e) 1601 MCTS, (h) 4501 MCTS, (k) 9001 MCTS. Type 2 CNV for $\rho_\text{adh}^{\Sigma_\sigma, \Sigma_{\sigma'}}$(RPE - BM) $ = 0$, snapshots at times: (c) 601 MCTS, (f) 1601 MCTS, (i) 4501 MCTS, (l) 9001 MCTS. We have represented the level curves of external VEGF as continuous lines. }
		\label{figAdh1} 
	\end{center} 
\end{figure}

Reducing the adhesion between the basement membrane of the RPE and BM may enable CNV to invade the sub-RPE space \cite{shi12, vamsi05}. As chemotaxis attracts vessel sprouts towards sites with higher VEGF concentration beyond the RPE, vessels may cross this layer at sites where adhesion is weakest, e.g., near drusen. If adhesion between RPE and BM is weak (large Potts parameter), ECs move easily in the space between them, thereby producing type 1 CNV, as observed in the left panels of Fig. \ref{figAdh1}. If the Potts parameter decreases (central and right panels of Fig.~\ref{figAdh1}), the adhesion between RPE and BM increases. ECs then try to surpass the RPE near the drusen that have opened a hole in BM. Eventually, the sprouts reach the subretinal space, whereby producing type 2 CNV. The  resulting CNV does not form a dense network of blood vessels between RPE and BM. 


\begin{figure}[h]
	\begin{center}
		\includegraphics[width=0.85\linewidth]{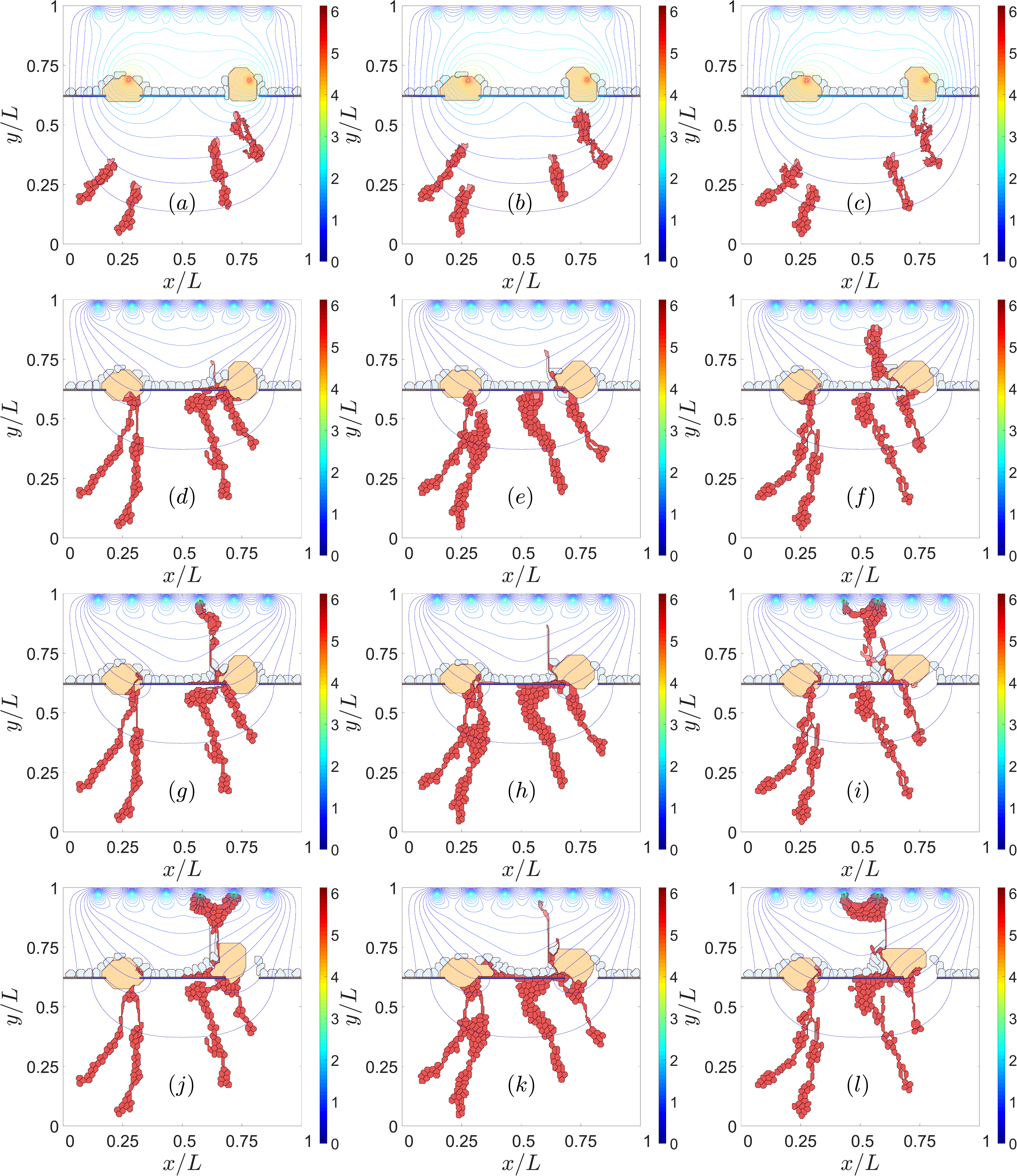}
		\caption{Effect of impaired adhesion between RPE - RPE \& EC - EC. Type 2 CNV for $\rho_\text{adh}^{\Sigma_\sigma,\Sigma_{\sigma'}}$(EC-EC) $ = 70$ and $\rho_\text{adh}^{\Sigma_\sigma,\Sigma_{\sigma'}}$(RPE cell - RPE cell) $ = 90$, snapshots at times: (a) 601 MCTS, (d) 1801 MCTS, (g) 3601 MCTS, (j) 8001 MCTS. Type 1 CNV for $\rho_\text{adh}^{\Sigma_\sigma, \Sigma_{\sigma'}}$(EC-EC) $ = 70$ and $\rho_\text{adh}^{\Sigma_\sigma,\Sigma_{\sigma'}}$(RPE cell - RPE cell) $ = 80$, snapshots at times: (b) 601 MCTS, (e) 1801 MCTS, (h) 3601 MCTS, (k) 8001 MCTS. Type 2 CNV for $\rho_\text{adh}^{\Sigma_\sigma, \Sigma_{\sigma'}}$(EC-EC) $ = 80$ and $\rho_\text{adh}^{\Sigma_\sigma,\Sigma_{\sigma'}}$(RPE cell - RPE cell) $ = 80$, snapshots at times: (c) 601 MCTS, (f) 1801 MCTS, (i) 3601 MCTS, (l) 8001 MCTS.}
		\label{figAdh2}
	\end{center} 
\end{figure}

\subsubsection{RPE - RPE \& EC - EC adhesion}
Impaired lateral adhesion between RPE cells facilitates type 2 CNV \cite{imamura05,simo10}. ECs and the sprouts they generate are able to penetrate the RPE layer effortlessly, as shown in the left column of Fig. \ref{figAdh2}. Stronger adhesion between cells in RPE makes it difficult for sprouts to cross the layer, thereby favoring type 1 over type 2 CNV, as shown in  the middle column of Fig. \ref{figAdh2}.

While adhesion between endothelial cells affects the quality of the resulting blood vessels \cite{ramos18}, it also influences the resulting type of CNV, cf the middle and right columns of Fig. \ref{figAdh2}. Reduced adhesion between ECs has the consequences displayed on the right column of Fig. \ref{figAdh2}: ECs are able to intersperse RPE cells and drusen to change quickly from type 1 to type 2 CNV. This produces blood vessels of poorer quality. Strong EC-EC adhesion makes it difficult for the sprout to pass through RPE cells since the ECs have to disconnect from their EC neighbors to cross the RPE, as shown by the middle column of Fig. \ref{figAdh2}. 

To sum up, Fig~ \ref{figAdh2} shows that we can favor type 1 CNV and prevent type 2 CNV by making stronger the adhesion between cells in RPE (from left column to middle column). In addition, if the adhesion between ECs weakens, poor quality sprouts will pass RPE layer and produce type 2 CNV (from middle column to right column).

\subsection{Sources of VEGF}
\label{subsec:VEGF}
\begin{figure}[h]
	\begin{center}		
		\includegraphics[width=0.85\linewidth]{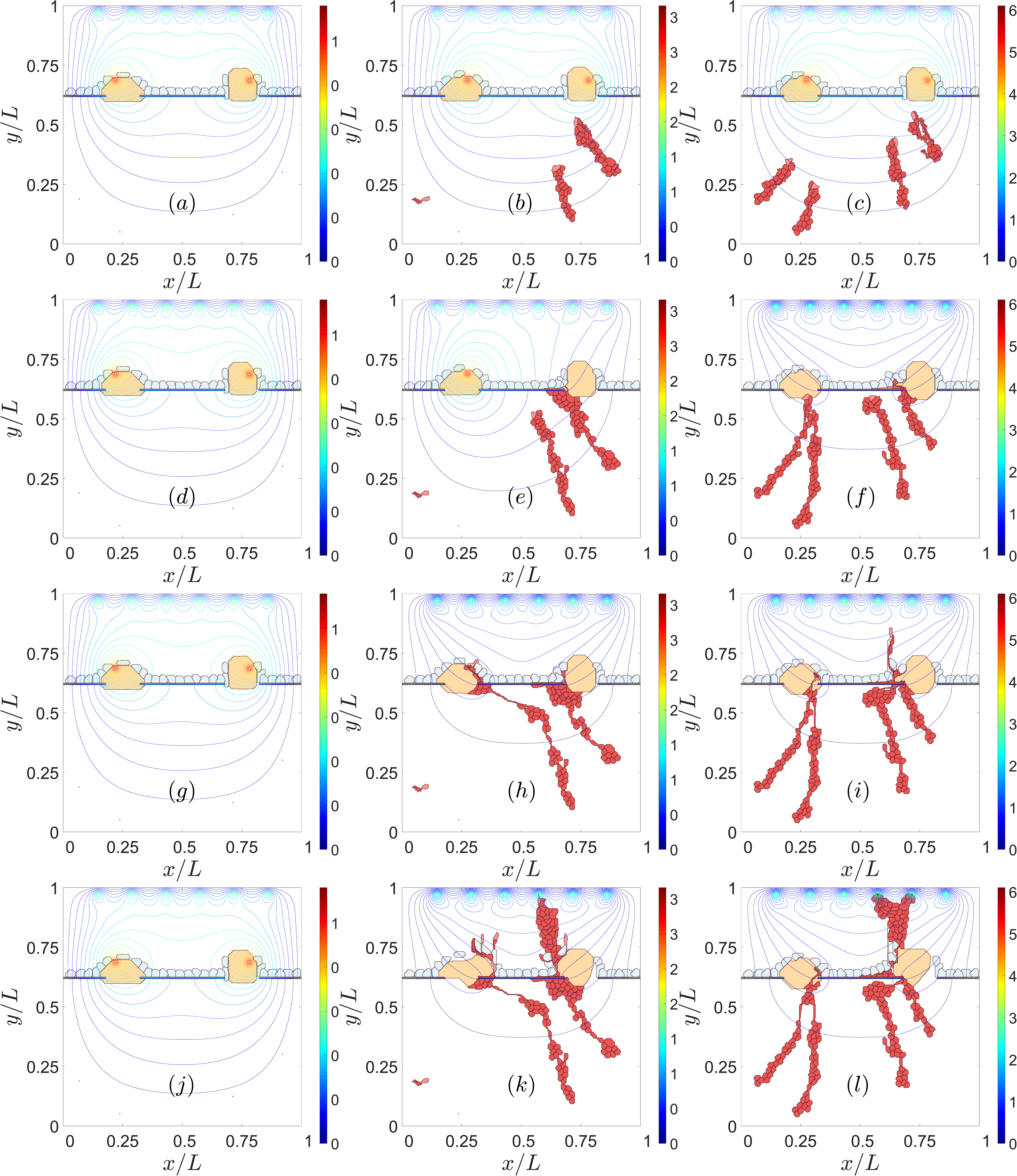}
		\caption{Effect of the VEGF concentration at the sources. For $\alpha_i = 4.01 \times 10^{-4} = 0.000401 $, snapshots at times: (a) 601 MCTS, (d) 1201 MCTS, (g) 3001 MCTS, (j) 5201 MCTS. For $\alpha_i  = 1.203 \times 10^{-3} =  0.001203$, snapshots at times: (b) 601 MCTS, (e) 1201 MCTS, (h) 3001 MCTS, (k) 5201 MCTS. For $\alpha_i = 2.005 \times 10^{-3}   = 0.002005$, snapshots at times: (c) 601 MCTS, (f) 1201 MCTS, (i) 3001 MCTS, (l) 5201 MCTS.}
		\label{figVEGF1}
	\end{center} 
\end{figure}

High levels of VEGF concentration generated by sources produce large VEGF gradients that drive sprouts, therefore being one chief cause of CNV \cite{bhutto06}. This is illustrated by Fig. \ref{figVEGF1}. The VEGF concentration at the sources in this figure increases from the left column to the middle and right ones, whereas time as measured by MCTS increases from top to bottom. If the level of VEGF is too low, the ECs at the walls of the choroid vessels do not have enough VEGF to activate and start to develop a sprout, as shown on the left column of Fig. \ref{figVEGF1}. Medium and high levels of VEGF concentration produce CNV, cf middle and right column of Fig.~\ref{figVEGF1}.  On the middle column of this figure, only two of the four possible choroid vessels that emit sprouts have been activated, whereas all four sites have been activated on the right column of Fig.~\ref{figVEGF1}. The subsequent larger chemotaxis causes the sprouts to reach the let drusen earlier on the right column of Fig.~\ref{figVEGF1} than on its middle column. The larger levels of VEGF favor the faster evolution from type 1 to type 2 CNV shown on the right column of Fig.~\ref{figVEGF1}.

\begin{figure}[h]
	\begin{center}
		\includegraphics[width=0.85\linewidth]{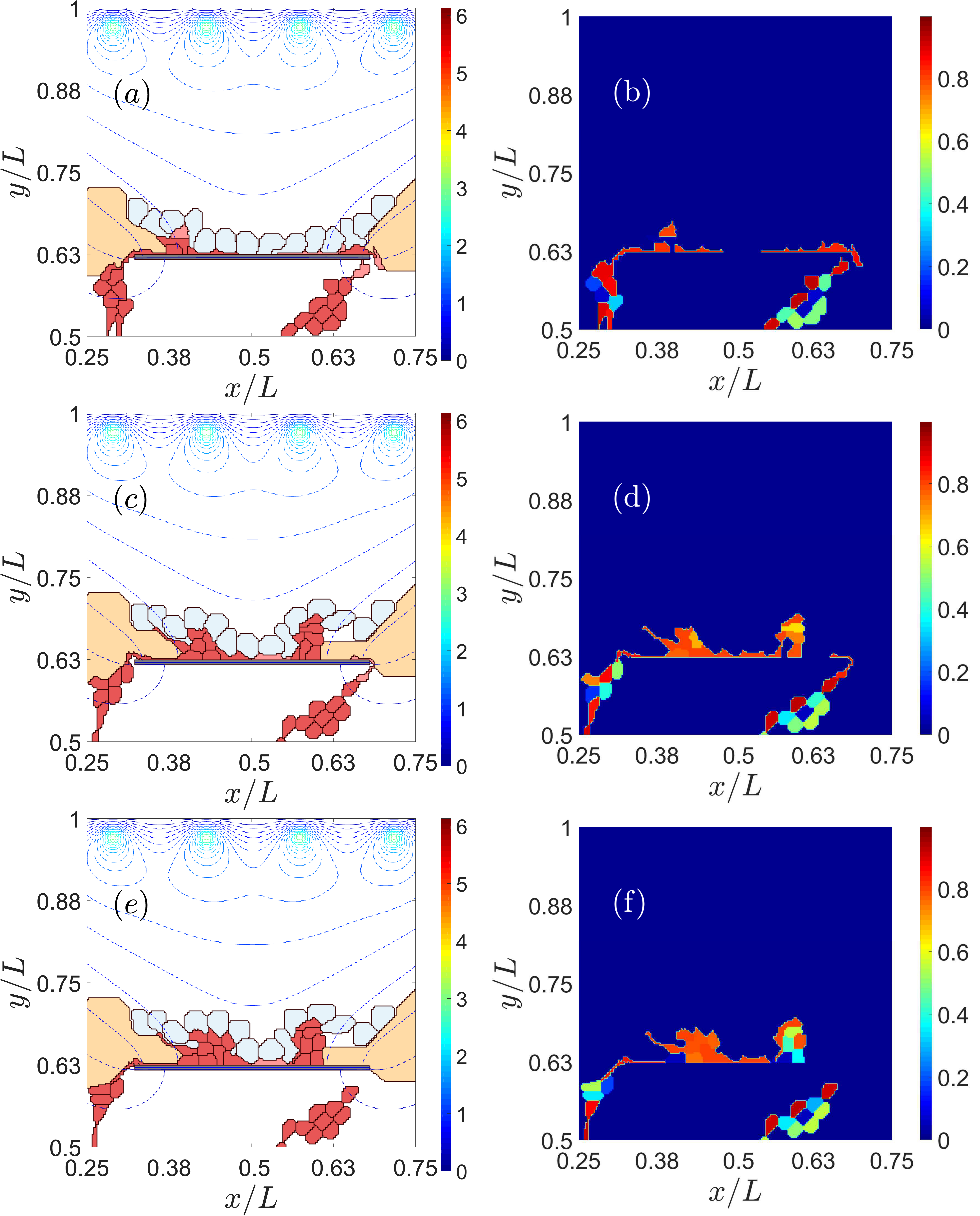}
		\caption{Type 1 CNV due to the lack of VEGF in the point where the tip cell of the sprout tries to cross the RPE (left column: (a), (c), (e)). Amount of VEGF receptors of ECs (right column: (b), (d), (f)). Snapshots at times: (a), (b) 1501 MCTS, (c), (d) 4501 MCTS, (e), (f) 9001 MCTS.}
		\label{figVEGF3}
	\end{center} 
\end{figure}

\begin{figure}[h]
	\begin{center}
		\includegraphics[width=0.85\linewidth]{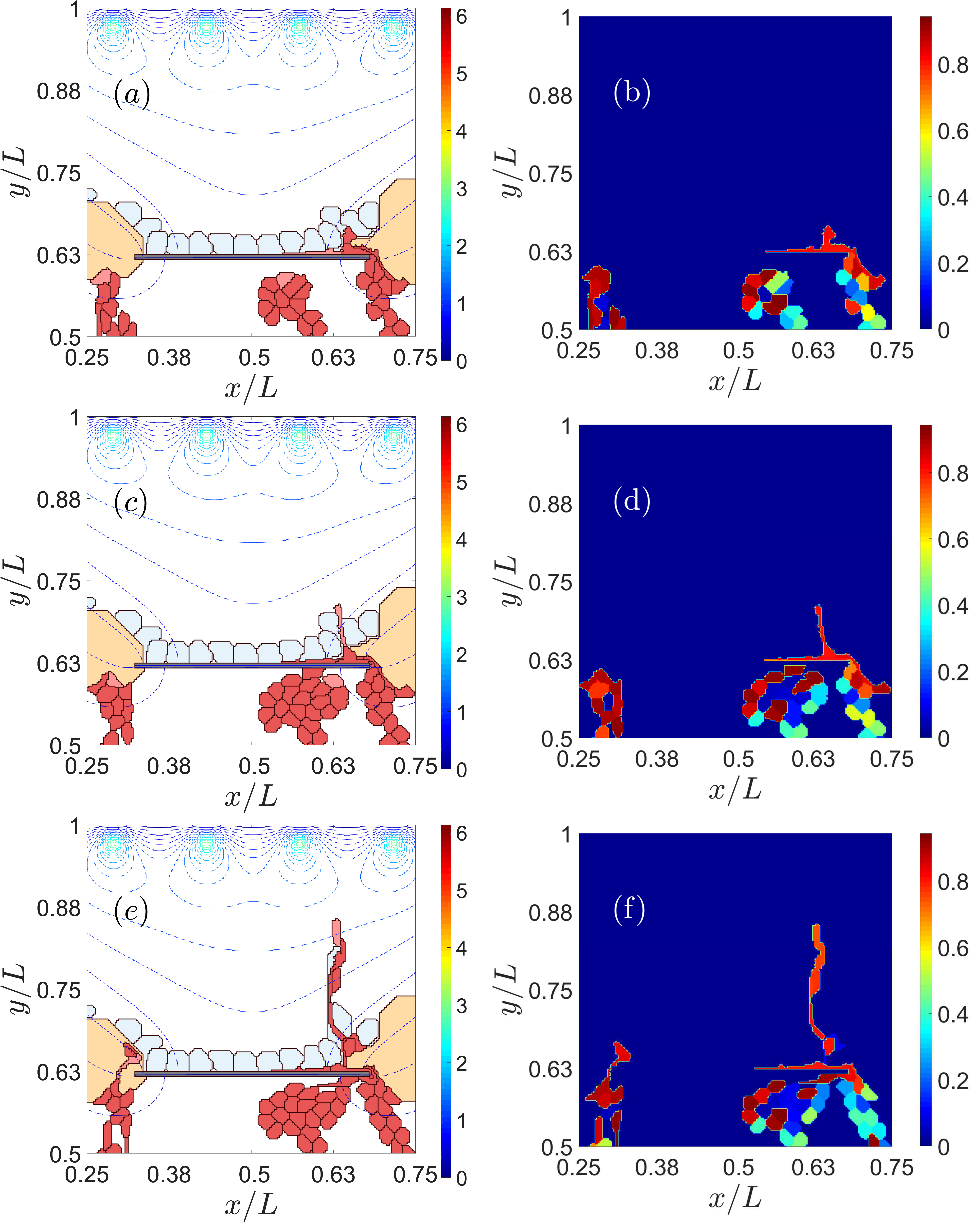}
		\caption{Type 2 CNV favored by the point where the sprout cross the RPE (left column: (a), (c), (e)). Amount of VEGF receptors of ECs (right column: (b), (d), (f)). Snapshots at times: (a), (b) 1201 MCTS, (c), (d) 1501 MCTS, (e), (f) 3001 MCTS.}
		\label{figVEGF4}
	\end{center} 
\end{figure}

The value of the VEGF gradient at the point where the sprout tries to cross the RPE determines the sprout chances of starting type 2 CNV. The VEGF concentration throughout the domain and the parameter values are the same for Figs.~\ref{figVEGF3} and \ref{figVEGF4}, which have a different seed of the random number generator that determines the sprout initiation points. The local VEGF gradient at the point where the sprouts are closest to the end of the RPE is larger for Fig.~\ref{figVEGF4} than for Fig.~\ref{figVEGF3}. The larger chemotactic force experienced by the leading EC implies that type 2 CNV is produced in Fig.~\ref{figVEGF4} while only type 1 CNV is observed in Fig.~\ref{figVEGF3}. The right columns of figures \ref{figVEGF3} and \ref{figVEGF4} depict the number of active VEGF receptors at the times corresponding to panels on the left columns. It is clear that the number of active VEGF receptors is larger when there is successful type 2 CNV, as in Fig. \ref{figVEGF4}, as compared with type 1 CNV as in Fig. \ref{figVEGF3}. Having the same adhesion and VEGF concentration do not determine the type of CNV. The setup of the parent vessels in the choroid may generate different CNV outcomes.


\begin{figure}[h]
	\begin{center}
		\includegraphics[width=0.85\linewidth]{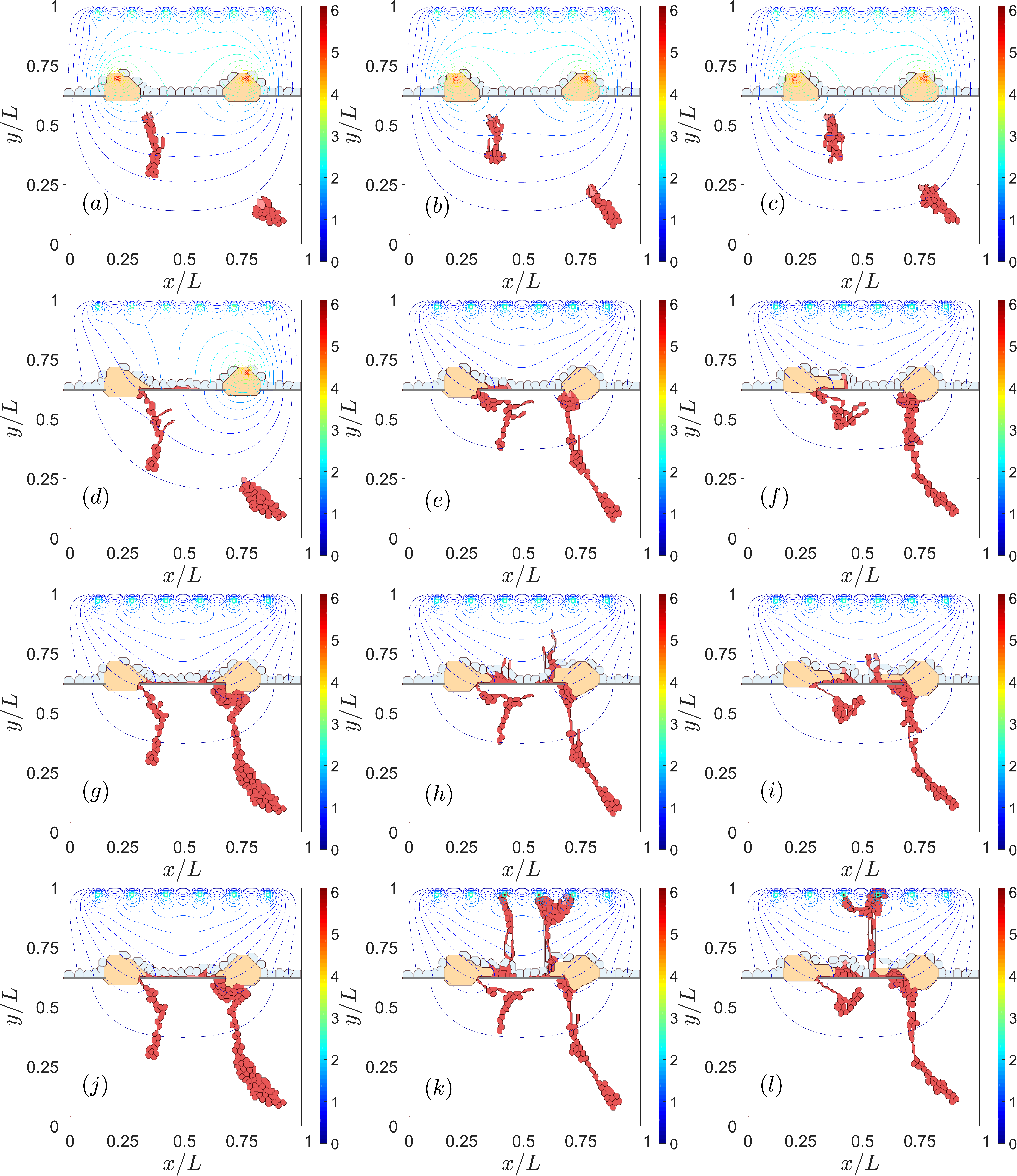}
		\caption{Effect of Jagged and Delta production on CNV. Type 1 CNV for $r_J = 500$ molec/h, $r_D = 1000$ molec/h, snapshots at times: (a) 601 MCTS, (d) 1601 MCTS, (g) 4501 MCTS, (j) 9001 MCTS. Type 2 CNV (reference simulation) for $r_J = 2000$ molec/h, $r_D = 1000$ molec/h, snapshots at times: (b) 601 MCTS, (e) 1601 MCTS, (h) 4501 MCTS, (k) 9001 MCTS. Type 2 CNV for $r_J = 2000$ molec/h, $r_D = 7500$ molec/h, snapshots at times: (c) 601 MCTS, (f) 1601 MCTS, (i) 4501 MCTS, (l) 9001 MCTS.}
		\label{figrJrD1} 
	\end{center} 
\end{figure}


\begin{figure}[h]
	\begin{center}
		\includegraphics[width=0.85\linewidth]{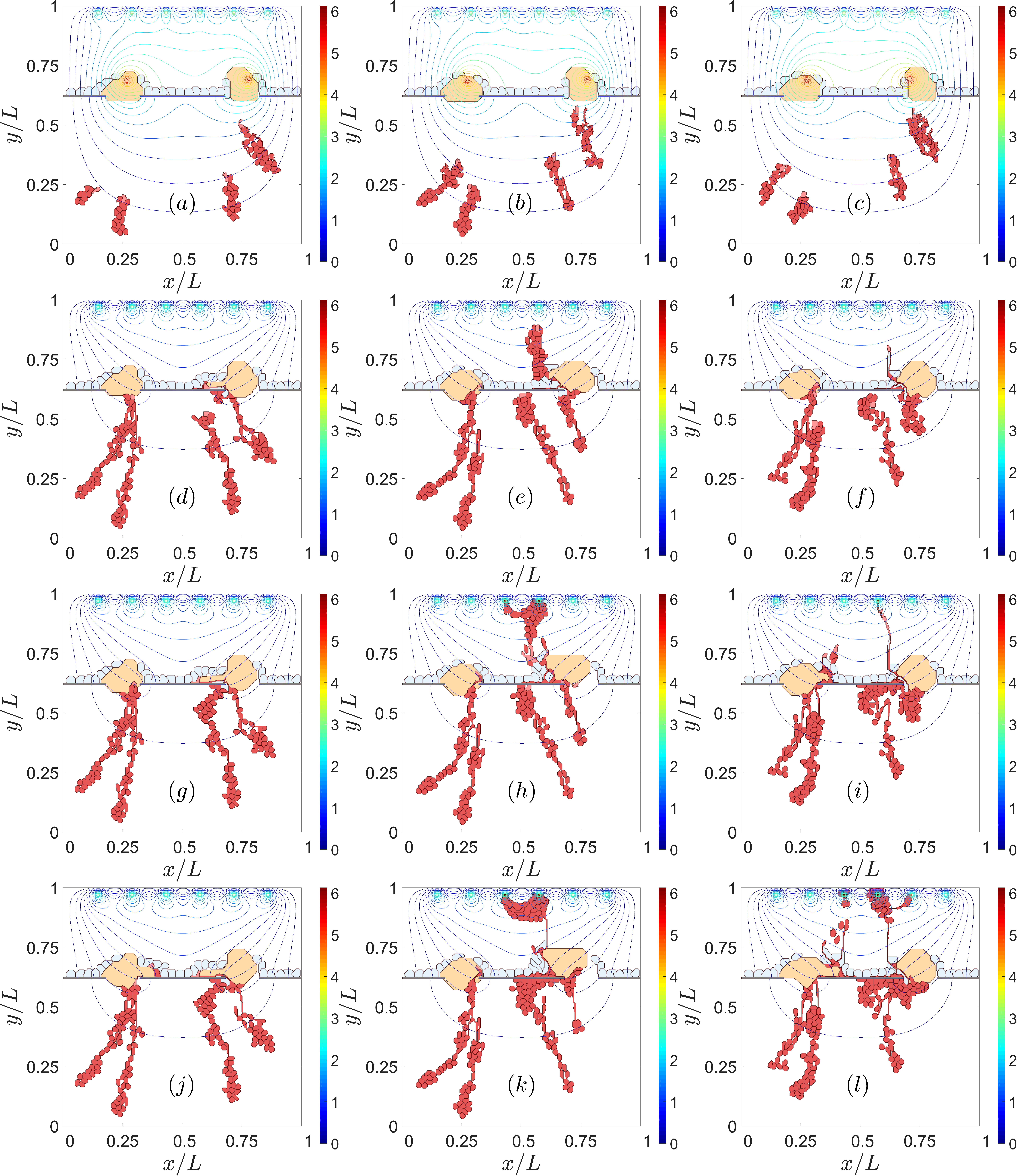}
		\caption{Effect of Jagged and Delta production on CNV. Type 1 CNV for $r_J = 500$ molec/h, $r_D = 1000$ molec/h, snapshots at times: (a) 601 MCTS, (d) 1801 MCTS, (g) 3601 MCTS, (j) 8001 MCTS. Type 2 CNV (reference simulation) for $r_J = 2000$ molec/h, $r_D = 1000$ molec/h, snapshots at times: (b) 601 MCTS, (e) 1801 MCTS, (h) 3601 MCTS, (k) 8001 MCTS. Type 2 CNV for $r_J = 2000$ molec/h, $r_D = 7500$ molec/h, snapshots at times: (c) 601 MCTS, (f) 1801 MCTS, (i) 3601 MCTS, (l) 8001 MCTS.
			}
		\label{figrJrD2}
	\end{center} 
\end{figure}

\subsection{Notch signaling}
\label{subsec:notch}
While Figs.~\ref{figVEGF3} and \ref{figVEGF4} show the effect of active VEGF receptors on retinal CNV, other proteins involved in the Notch signaling pathway may characterize the resulting CNV network.

Jagged and Delta dynamics determine sprouting \cite{veg20}, therefore also CNV. The thresholds of Delta concentration of cells to change the phenotype depend on the Jagged production rate $r_J$. The simulations shown previously have a Jagged production rate, $r_J=2000$, and a Delta production rate, $r_D=1000$. The chosen values correspond to pathological angiogenesis \cite{veg20}, which is the most similar scenario to type 2 CNV.
Figs.~\ref{figrJrD1} and \ref{figrJrD2} show the effect of decreasing $r_J$ in the left column and increasing $r_D$ in the right column from the reference simulation placed in the middle of the figures. The reference simulation in Fig.~\ref{figrJrD1} is the one corresponding to middle column in Fig.~\ref{figAdh1}. The reference simulation in Fig. \ref{figrJrD2} is the one corresponding to right column in Fig. \ref{figAdh2}. In both figures, \ref{figrJrD1} and \ref{figrJrD2}, reducing the Jagged production rate, left column, favors type 1 over type 2 CNV and makes the blood vessels thicker than the ones in the middle column. However, the increase of the Delta production rate does not prevent type 2 CNV. We also observe that the resulting blood vessels are thinner and worse organized than the ones in the middle column.

\section{Discussion}\label{sec:discussion}  
The mathematical model of angiogenesis in retina presented here illustrate the relative importance of mechanical, chemical and cellular cues to study AMD. When Bruch's membrane deteriorates, transport of oxygen, nutrients and debris across it become impaired and there appear drusen deposits that displace retinal pigment epithelial cells. In our model, the drusen deposits further deteriorate BM and affect the RPE cells. The latter may undergo changes in their adhesion properties, experience hypoxia and become local sources of VEGF (we do not distinguish here different varieties of VEGF). The extra VEGF reaches choroid vessels, which may issue sprouts led by tip endothelial cells if the local VEGF concentration surpasses a threshold. These sprouts form the choroidal neovascularization, whose type depends on the impaired adhesion between BM and RPE and between the RPE cells themselves. Notch signaling decides the EC phenotype, and alterations thereof strongly influence CNV type and configuration. In particular, overexpression of Jagged as shown by a large production thereof yields high proliferation of stalk ECs and thin vessels led by hybrid type tip cells. These leaky vessels could be important in exudative AMD.

Let us now examine in some detail the roles of adhesion defects and Notch signaling in CNV. Impaired adhesion between the basement membrane of the RPE and BM allow the cells to move easily in this space, which may produce type 1 CNV. If the adhesion between RPE and BM is strong, vessel sprouts may not be able to move between them, which impedes type 1 CNV. Our simulations show that a reduced lateral adhesion between RPE cells makes it easier for ECs to cross the RPE to the subretinal space, thereby producing type 2 CNV. Even if the impaired adhesion between BM and RPE allows formation of type 1 CNV, weaker lateral adhesion between RPE cells may facilitate  vessel sprout crossing of the RPE layer. Then, the CNV changes from type 1 to type 2. 

Drusen deposits pinpoint the deterioration of the RPE and BM, which are the hallmarks of dry AMD and may produce exudative (wet) AMD. In our model, extra sources of VEGF appear near drusen and trigger angiogenesis. Large VEGF gradients result and may cause CNV. While a low level of VEGF at the walls of the choroid vessels does not activate ECs that start a sprout, medium and high VEGF levels produce CNV and the number of sprouts that are activated depends on these levels. The VEGF concentration is not only related to CNV start, but also to its type. Under the same parameter values and conditions, the value of the VEGF gradient at the point where the sprout tries to cross the RPE determines the sprout chances of starting type 2 CNV. Higher VEGF concentration produces more active VEGF and greater number of VEGF receptors in tip cells so that the larger chemotactic force experienced by the leading ECs favors type 2 CNV over type 1 CNV.

In the Notch signaling pathway, VEGF receptors, active VEGF and a large Jagged production rate produce pathological angiogenesis \cite{veg20}, which is most similar to type 2 CNV. Decreasing the Jagged production rate yields thicker blood vessels  and reduces the number of hybrid cells behaving as tip cells, thereby bringing in type 1 CNV. While a large Delta production rate counters a high Jagged production rate \cite{veg20}, it does not prevent type 2 CNV in retina. In this case, the resulting blood vessels are thinner and worse organized, but they are still able to cross the RPE layer, as shown in Figs.~\ref{figrJrD2}{\em(c,f,i,l)}. The content of VEGF in the ECs as a function of the external VEGF or of the Jagged content also indicates the CNV type. The wider the range of VEGF content is, the greater the probability of having type 2 CNV is. 

Our simulations suggest possible therapies based on using drugs to tinker with parameters controling CNV outcomes. Anti-VEGF treatments are commonly used to stop angiogenesis and provide a temporary reprieve of exudative AMD \cite{fog18}. Our simulations also corroborate that lowering VEGF production stops angiogenesis. According to our results, anti-Jagged antibodies \cite{patent,sie17} could have a similar effect inhibiting CNV in AMD. To this end, progress in robotic-assisted subretinal injections may provide a beneficial and standardized implementation of anti-angiogenic therapies \cite{zho20}. The same end can be achieved by enforcing the adhesion properties of RPE cells to BM and among themselves. Promising experiments involve inhibition of the EMT transition to regenerate the RPE \cite{rad15}. Surgical procedures include RPE and choroid transplantation when appropriate donors are available \cite{par20}. Current lines of ongoing experimental preclinical research aim at placing implants to stimulate RPE and photoreceptors or replace their function \cite{che17,may17,tan18}, seed pluripotent stem cells to replace the lost RPE cells and photoreceptors \cite{gag19,sin20}, use gene addition therapies and genome editing to restore function to a non-functional or absent protein in the appropriate RPE signaling pathways, or to block function by knocking-down proteins \cite{sti20}, or repairing BM with biomaterials and growing RPE and photoreceptors over it \cite{jem20}. Clearly, the lifetime of working implants will depend on that of their underlying devices whereas transplants and stem cells may recover RPE cells and improve their lateral and BM adhesion properties, and also recover photoreceptors. However, BM may continue its defective performance and AMD may reappear with time. Replacing BM, RPE cells and photoreceptors is a more drastic but perhaps longer lasting solution if it works.  

To allow for quantitative comparisons with experiments, our 2D model of angiogenesis in retina needs to be extended in several directions to be made more realistic and to predict the evolution of wet AMD. The extension of the model to three dimensional configurations is straightforward although it requires more computing power. To move toward later stages of the formation of an advancing vascular plexus, we need to add lumen formation and blood circulation to the model. These processes will allow us to tackle the concurrent sprouting and anastomosis on the front of the advancing vascular plexus and the pruning of poorly perfused sprouts on its back. 

\section{Conclusions} \label{sec:conclusions}
To conclude, we have proposed a cellular Potts model of wet AMD that accounts for deterioration of Bruch's membrane, growing drusen deposits that turn on sources of VEGF in addition to those active near photoreceptors. Numerical simulations of the model show that choroid neovascularization mainly results from three causes: (i) impairment  of the adhesion between retinal pigmentation epithelium cells, between these cells and Bruch's membrane and among endothelial cells; (ii) excess VEGF producing strong gradients thereof, (iii) excess Jagged production. Anti-VEGF and anti-Jagged treatments address (ii) and (iii) and could halt angiogenesis on a temporary basis, but do they not resolve the deterioration of Bruch's membrane that produces AMD. While anti-VEGF treatments are standard for wet AMD \cite{fog18}, anti-Jagged drugs have been tried in cancer treatments and more research would be useful to ascertain their value for AMD \cite{patent,sie17}. Our numerical simulations suggest the need for further experiments to confirm our findings, sharpen and validate our AMD model. More drastic remedies to address cellular adhesion loss (i) are electrobiological implants replacing the function of RPE and photoreceptors \cite{che17,may17,tan18}, seeding pluripotential stem cells to replace the lost RPE and photoreceptor cells themselves \cite{gag19,sin20}, or replacing  Bruch's membrane by an artificial biohybrid retina \cite{jem20}. These researches are still in early preclinical stages and are much more intrusive for aged patients. Experiments to quantify the adhesion of the new cells generated by pluripotent stem cells and their ability to stop angiogenesis would be needed. Our passive model of RPE cells could be replaced by a vertex model of the epithelium able to describe wound healing \cite{bon20}. Including reversible EMT signaling pathways \cite{shu20} in future models would be desirable and could bring about new therapies. Modeling and numerical simulation can thus be key to identifying critical experiments that are most likely to improve our understanding of AMD and possible treatments.
\bigskip


\noindent {\bf Acknowledgments.} This research was funded by FEDER/Ministerio de Ciencia, Innovaci\'on y Universidades -- Agencia Estatal de Investigaci\'on grant number MTM2017-84446-C2-2-R. 

\appendix
\setcounter{equation}{0}
\renewcommand{\theequation}{A.\arabic{equation}}
\setcounter{table}{0}
\renewcommand{\thetable}{A.\arabic{table}}
\section{Appendix A: Cellular Potts model}\label{app1}
We consider a square domain $\Omega$ of side $L$ (in numerical simulations, $L= 230\ \mu$m) having $M$ nodes on each side. On $\Omega$, there are $M\times M$ grid points and $(M-1)^2$ elementary squares (pixels) $\mathbf{x}$, each having an area $L^2/(M-1)^2$. Pixels $\mathbf{x}$ can belong to different cells or ECM, $\Sigma_\sigma$, and are labelled by their spin $\sigma(\mathbf{x})$ as follows: 
\begin{equation}
\sigma(\mathbf{x}) = \left\lbrace
\begin{array}{cc}
0, & \text{ if } \mathbf{x} \in \Sigma_\text{ECM}; \\
1, \dots, \ N_{\text{drusen}} + 1 , & \text{ if } \mathbf{x} \in \Sigma_\text{BM}; \\
N_{\text{drusen}} + 2  \dots, \ 2N_{\text{drusen}} + 1 , & \text{ if } \mathbf{x} \in \Sigma_\text{drusen}; \\
2N_{\text{drusen}} + 2  \dots, \ 2N_{\text{drusen}} + 1 + N_{\text{RPE}}, & \text{ if } \mathbf{x} \in \Sigma_\text{RPE}; \\
2N_{\text{drusen}} + N_{\text{RPE}} + 2, \dots  & \text{ if } \mathbf{x} \in \Sigma_\text{EC}.
\end{array}\right.\label{a1}
\end{equation}
where $N_{\text{drusen}}$ is the number of drusen and $N_{\text{RPE}}$ is the number of RPE cells.

\paragraph{Energy functional.} In the Hamiltonian of Eq.~\eqref{eq1}, we have
\begin{itemize}
\item The net variation of the durotaxis term $H_{\text{durot}}$ is \cite{oer14}
\begin{eqnarray}
\Delta H_{\text{durot}}\! =\! - \rho_{\text{durot}}\, g(\mathbf{x},\mathbf{x'})\, f(\mathbf{x},\mathbf{x'}), \label{a4}
\end{eqnarray}
where $\rho_{\text{durot}}$ is a Potts parameter, $g(\mathbf{x},\mathbf{x'}) = 1$ for extensions and $g(\mathbf{x},\mathbf{x'}) = -1$ for retractions, and the function $f(\mathbf{x},\mathbf{x'})$ depends on the solution of the elasticity equations as described in \cite{veg20}. 

\item The variation of the chemotaxis term $H_{\text{chem}}$ is \cite{bau07}
\begin{equation}
\Delta H_{\text{chem}} = - \rho_{\text{chem}}(\mathbf{x},\mathbf{x'}) \frac{C(\mathbf{x'}) - C(\mathbf{x})}{1 + 0.3\, C(\mathbf{x})}, \label{a5}
\end{equation}
where $C$ is the VEGF concentration in the corresponding pixel, given by Eqs.~\eqref{eq2}, and $\rho_{\text{chem}}(\mathbf{x},\mathbf{x'})\geq 0$ is a Potts parameter that depends on the type of EC or ECM occupying pixels $\mathbf{x}$ and $\mathbf{x'}$. We have 
\begin{eqnarray}
\rho_\text{chem}(\mathbf{x},\mathbf{x'})=\frac{\rho_\text{chem}^0}{\max_k D_k}\left\{\begin{array}{cc}
D_i, & \mathbf{x}\in\Sigma_i,\, \mathbf{x'}\in\Sigma_\text{ECM} \mbox{ or vice versa,}\\
\frac{D_i+D_j}{2}, & \mathbf{x}\in\Sigma_i,\,\mathbf{x'}\in\Sigma_j \mbox{ or vice versa,}\\
\end{array}\right.   \label{a6}
\end{eqnarray}
where $i$ and $j$ are ECs. The positive constant $\rho_\text{chem}^0$ measures the magnitude of chemotaxis. The level of Delta-4 determines the EC phenotype and, according to Eqs.~\eqref{a5} and \eqref{a6}, the strength of their chemotactic drive. Tip cells have a higher level of Delta-4 and, consequently by Eq.~\eqref{a6}, they are more motile than stalk cells. 
\end{itemize}

\begin{table}[h]
	\begin{tabular}{r|c|c|c|c|c|c|c|c|c}
		\text{Parameter} & $\rho_\text{area}${\scriptsize (EC)} & $\rho_\text{area}${\scriptsize (RPE)}  & $\rho_\text{area}${\scriptsize (DR)} &  $\rho_\text{perim}${\scriptsize (EC)} &  $\rho_\text{perim}${\scriptsize (RPE)} &  $\rho_\text{perim}${\scriptsize(DR) }&  $\rho_\text{length}${\scriptsize(EC)} & $\rho_\text{durot}$ & $\rho_\text{chem}^0$\\ \hline
		\text{Value} &  25000 & 100000 & 750000  & 75 & 100 & 500  &180&25&50000
	\end{tabular}
	\caption{Dimensionless Potts parameters corresponding to the area, perimeter, length, durotaxis and chemotaxis constraints (DR=drusen) \cite{veg20}.}
	\label{t1}
\end{table}

\begin{table}[h]
	\begin{tabular}{r|c|c|c|c|c|c|c|c|c|c}
		\text{Parameter $\rho_\text{adh}^{\Sigma_\sigma,\Sigma_{\sigma'}}$} &{\scriptsize EC-EC} &{\scriptsize EC-ECM} &{\scriptsize EC-RPE}&{\scriptsize EC-DR} & {\scriptsize RPE-ECM}  & {\scriptsize RPE-BM} & {\scriptsize RPE-DR}  & {\scriptsize RPE-RPE } & {\scriptsize DR-ECM} & {\scriptsize DR-DR} \\ \hline
		\text{Value} & 70-80&40&60&80&40 & 0-30 & 160 & 80-90 & 80 &200
	\end{tabular}
	\caption{Dimensionless Potts parameters $\rho_\text{adh}^{\Sigma_\sigma,\Sigma_{\sigma'}}$ for adhesion (DR=drusen) \cite{veg20}.}
	\label{t2}
\end{table}

The values of the Potts parameters are listed in Tables \ref{t1} and \ref{t2} from Ref.~\cite{veg20}. They are adjusted so that the terms in the net variation of the hamiltonian all have the same order. The perimeter contribution, absent in Refs.~\cite{bau07,oer14}, is small compared to the other terms in Eq.~\eqref{eq1}, so that it only affects the computations in extreme cases (e.g., extremely thin cells, thin cells that stick to the blood vessel). The sensitivity of this CPM to parameter values is discussed in Ref.~\cite{veg20}.

\setcounter{equation}{0}
\renewcommand{\theequation}{B.\arabic{equation}}
\setcounter{table}{0}
\renewcommand{\thetable}{B.\arabic{table}}
\section{Appendix B: Notch signaling pathway equations} \label{app3}
 The Notch signaling pathway is activated when the transmembrane receptor Notch belonging to a particular cell interacts with the transmembrane ligands Delta-4 or Jagged-1 belonging to its neighboring cell (trans-activation), thereby releasing the Notch intracellular domain (NICD). NICD then enters the nucleus and modulates the expression of many target genes of the Notch pathway, including both the ligands Delta and Jagged. However, when Notch of a cell interacts with Delta or Jagged belonging to the same cell, no NICD is produced; rather, both the receptor (Notch) and ligand (Delta or Jagged) are degraded (cis-inhibition) and therefore the signaling is not activated. These mechanisms are incorporated into model differential equations proposed in Ref.~\cite{boa15}, and coupled to the CPM in Ref.~\cite{veg20}. Here, we describe these equations for a given cell $i$ surrounded by other cells. At time $t$, let $N_i$, $D_i$, and $J_i$ be the number of Notch, Delta-4, and Jagged-1 proteins in the $i$th cell, respectively. Similarly, let $I_i$, $V_{Ri}$ and $V_i$ be the number of NICD, VEGF receptor and VEGF molecules, respectively. These variables satisfy the following equations, 
\begin{subequations}\label{c1}
\begin{eqnarray}
&&\frac{dN_i}{dt}  =  r_N H^{S}(I_i, \lambda_{I,N}) - \{ [k_C D_i + k_T D_\text{ext}(i)]\, H^{S}(I_i,\lambda_{D,F}) +\gamma \nonumber\\
&&\quad\quad+\, [k_C J_i + k_T J_\text{ext}(i)]\, H^{S}(I_i, \lambda_{J,F}) \} \, N_i, \label{c1a}\\
&&\frac{dD_i}{dt}  =  r_D H^{S}(I_i, \lambda_{I,D}) H^{S}(V_i, \lambda_{V,D}) - \!\left[k_C N_i  H^{S}(I_i, \lambda_{D,F}) + k_T N_\text{ext}(i)+ \gamma\right] D_i, \label{c1b}\\
&&\frac{dJ_i}{dt}  =  r_J H^{S}(I_i,\lambda_{I,J}) - \!\left[k_C N_i  H^{S}(I_i,\lambda_{J,F}) + k_T N_\text{ext}(i) + \gamma\right] J_i, \label{c1c}\\
&&\frac{dI_i}{dt}  =  k_T N_i \!\left[ H^{S}(I_i,\lambda_{D,F}) D_\text{ext}(i) + H^{S}(I_i,\lambda_{J,F}) J_\text{ext}(i)\right]\! - \gamma_S I_i, \label{c1d}\\
&&\frac{dV_{Ri}}{dt}  = r_{VR} H^{S}(I_i,\lambda_{I,V_R}) -  k_T V_{Ri} V_\text{ext}(i) - \gamma V_{Ri}, \label{c1e}\\
&&\frac{dV_i}{dt} =   k_T V_{Ri} V_\text{ext}(i) - \gamma_S V_i.  \label{c1f}
\end{eqnarray}
\end{subequations}
Here, $r_N$, $r_D$, $r_J$, and $r_{VR}$, are the production rates of $N$, $D$, $J$, and $V_R$, respectively. The cis-inhibition and trans-activation rates are $k_C$ and $k_T$, respectively, whereas $\gamma$ and $\gamma_S$ are degradation rates for $N$, $D$, $J$, $V_R$ and for $I$, $V$, respectively. These parameters, their representative values and units are listed in Table \ref{t3}. The shifted, excitatory and inhibitory Hill functions appearing in Eqs.~\eqref{c1} are:
\begin{subequations}\label{c2}
\begin{eqnarray}
&&H^S(\xi, \lambda_{\eta,\zeta}) = H^-(\xi) + \lambda_{\eta,\zeta} H^+(\xi), \label{c2a}\\
&&H^-(\xi) = \frac{1}{1 + \left(\frac{\xi}{\xi_0}\right)^{n_\zeta}}, \quad H^+(\xi)  = 1 - H^-(\xi),\label{c2b}
\end{eqnarray}
\end{subequations}
where $H^{S}$ is excitatory for $\lambda_{\eta,\zeta} >1$ and inhibitory for $\lambda_{\eta,\zeta} \leq1$. In Eqs.~\eqref{c2}, $\xi=V, I$, $\eta=I,V,D,J$, and $\zeta=N,D,J,V_R,F$ (the subscript $F$ refers to Fringe, cf.~\cite{boa15}). The dimensionless parameters $n_\zeta$ and $\lambda_{\eta,\zeta}$ appearing in the Hill functions are listed in Table \ref{t4}. We solved Eqs.~\eqref{c1} with zero initial conditions for all unknowns but the outcome of the simulations does not change if other initial conditions are used.

\begin{table}[ht]
\begin{tabular}{r|c|c|c|c|c|c}
	\text{Parameter} & $r_N$    & $r_D$, $r_J$, $r_{VR}$ & $k_C$   & $k_T$  & $\gamma$ & $\gamma_S$   \\ \hline
	\text{Value}     & 1200    & 1000    & $5\times 10^{-4}$    & $2.5\times 10^{-5}$   & 0.1 & 0.5   \\ \hline
	\text{Unit}     & molec/h & molec/h & (h\, molec)$^{-1}$ & (h\, molec)$^{-1}$ & h$^{-1}$   & h$^{-1}$
\end{tabular}
\caption{Rates appearing in Eqs.~\eqref{c1}.}
\label{t3}
\end{table}

\begin{table}[ht]
\begin{tabular}{r|c|c|c|c|c|c|c|c|c}
	\text{Parameter} & $\lambda_{I,N}$, $\lambda_{V,D}$, $\lambda_{I,J}$   & $\lambda_{I,D}$,  $\lambda_{I,V_R}$  & $\lambda_{D,F}$ &  $\lambda_{J,F}$& $n_N$, $n_D$, $n_V$, $n_{V_R}$& $n_J$ & $n_F$& $I_0$, $V_0$& $\chi_V$\\ \hline
	\text{Value}     & 2.0    & 0.0  & 3.0 & 0.3 & 2.0 & 5.0 & 1.0 & 200& 1.0
\end{tabular}
\caption{Dimensionless parameters appearing in the Hill functions. $I_0$ and $V_0$ are activation numbers of NICD and VEGF molecules, respectively, and $\chi_V$ is the conversion factor.}
\label{t4}
\end{table}

If $j\in\langle i\rangle$ are the cells $j$ sharing boundary of length $P_{i,j}$ with cell $i$, the number of $X=N, D, J$ molecules outside cell $i$ is
\begin{eqnarray}
X_\text{ext}(i)=\frac{1}{P_i}\sum_{j\in\langle i\rangle}P_{i,j} X_j. \label{c3}
\end{eqnarray}
The perimeter of cell $i$, $P_i$, minus $\sum_{j\in\langle i\rangle}P_{i,j}$ is the length of its boundary that is not shared with any other cell. Note that $X_\text{ext}(i)$ is simply the sum of all $X_j$ if the lengths $P_{i,j}$ are all equal and $P_i=\sum_{j\in\langle i\rangle}P_{i,j}$ because the whole boundary of cell $i$ is shared with other cells. As the cell moves and its boundaries fluctuate due to cellular Potts dynamics, the membrane protein levels of the neighboring cells interacting with the moving cell also vary. In this way, the production rates of the different proteins in a cell are directly influenced by the interactions with its neighborhood and, in particular, by the membrane fluctuations of the cell. $V_\text{ext}(i)$ is the number of VEGF molecules outside the $i$th cell that interact with VEGF receptor cells to produce VEGF molecules inside the $i$th cell. The external VEGF cells come from the continuum field $C(x,y,t)$, which diffuses from $x=L$. Let $\mathbf{x}_i$ be the pixel of the $i$th cell that is closer to the hypoxic region. The number of external VEGF molecules in that pixel is $C(\mathbf{x}_i,t)$ multiplied by the conversion factor $\chi_V=N_AL^2/[(M-1)^2M_V]$, where $M_V$ is the molecular weight of the VEGF molecules and $N_A$ is the Avogadro number. We have used $\chi_V=1$, which is representative of VEGF molecules with a large molecular weight. In the numerical simulation, $C$ is known in the grid points and its value at a pixel should be the average value of the four grid points of the pixel. Since these values are quite similar, we adopt the value of $C$ at the bottom left grid point of the pixel $\mathbf{x}_i$ as $C(\mathbf{x}_i,t)$.

\begin{table}[ht]
	\begin{tabular}{r|c|c|c|c|c|c}
		\text{Variable} & $N_i$, $D_i$, $J_i$, $N_\text{ext}$, $D_\text{ext}$, $J_\text{ext}$  & $I_i$  & $V_{Ri}$ &  $V_i$ & $V_\text{ext}$ & $t$ \\ \hline
		\text{Scale}   & $\sqrt{r_D/k_C}$    & $(k_T r_D)/(k_C \gamma_S)$  & $r_{V_R}/ \gamma$ & $V_0$ & $6 V_0$ & $ 1/\sqrt{k_C r_D} $\\ \hline
		\text{Value}   & $ \sqrt{2} \times 10^3$ & $10^2$  & $10^4$ & $2\times 10^2$ & $12\times 10^2$ & $\sqrt{2}$\\ \hline
		\text{Unit}   & molec & molec & molec & molec & molec   & h
	\end{tabular}
	\caption{Units for nondimensionalizing the Notch equations \eqref{c1}.}
	\label{t5}
\end{table}


\end{document}